\documentclass[aps, prd, superscriptaddress,a4paper,twocolumn]{revtex4-1}
\usepackage{hyperref}
\usepackage{graphicx}
\usepackage[sort&compress]{natbib}
\usepackage{latexsym}
\usepackage{epsfig}
\usepackage[english]{babel}
\usepackage{graphicx}
\usepackage[T1]{fontenc}
\usepackage[utf8]{inputenc}
\usepackage{amsmath}
\usepackage{slashed,color}

\usepackage{comment}
\usepackage{natbib}
\usepackage{bm}
\usepackage{amssymb}

\usepackage{mathtools}

\date{\today}

\begin{document}
\title{The Chiral Phase Transitions of Helical Matter}

\author{Shen-Song Wan}
\affiliation{School  of  Nuclear  Science  and  Technology,  Lanzhou  University,
222  South  Tianshui  Road,  Lanzhou  730000,  China.}

\author{Marco Ruggieri}\email{ruggieri@lzu.edu.cn}
\affiliation{School  of  Nuclear  Science  and  Technology,  Lanzhou  University,
222  South  Tianshui  Road,  Lanzhou  730000,  China.}

\begin{abstract}
We study the thermodynamics of helical matter, namely quark matter 
in which a net helicity, $n_H$, is in equilibrium. 
Interactions are modeled by the renormalized quark-meson model with two flavors of quarks.
Helical density is described within the grand-canonical ensemble formalism via a chemical potential, $\mu_H$.
We study the transitions from the normal quark matter and hadron gas 
to the helical matter, drawing the phase diagram at zero temperature. 
We study the restoration of chiral symmetry at finite temperature and 
show that the net helical density softens the transition, moving the critical endpoint to lower temperature and higher
baryon chemical potential.
Finally, we discuss briefly the effect of a rigid rotation on the helical matter,
in particular on the fluctuations of $n_H$, and show that these are enhanced by the rotation. 
\end{abstract}
\pacs{.s}
\maketitle

\section{Introduction}

The phase diagram of Quantum Chromodynamics (QCD) 
in the temperature, $T$, and baryon chemical potential, $\mu$, plane
is one of the most active research subject in modern high energy physics.
At $\mu=0$,  first principles Lattice QCD calculations show that
QCD matter experiences a  crossover from
a low temperature confined phase, in which chiral symmetry is spontaneously broken,
to a high temperature phase in which chiral symmetry is approximately
restored \cite{Borsanyi:2013bia,Bazavov:2011nk,Cheng:2009zi,Borsanyi:2010cj,Borsanyi:2010bp}.
It is not possible to define
uniquely a critical temperature for the restoration of chiral symmetry: 
it is more appropriate to define a pseudo-critical region centered on a pseudo-critical temperature, $T_c$,
that is a range of temperatures in which several physical quantities  experience substantial changes.
The (pseudo)-critical temperature is found to be $T_c\approx 150$ MeV $\approx 10^{12}$ K.

Despite the amazing results obtained by Lattice QCD in recent years, 
it is not possible to perform reliable simulations of QCD with three colors at  large  $\mu$.
For this reason, the use of effective models for studying the phase transitions of QCD at $\mu\neq 0$ 
is necessary. In this context  
the quark-meson (QM) model is very popular
due to its renormalizability 
\cite{Skokov:2010sf,Zhang:2019neb,Frasca:2011zn,Ruggieri:2013cya,Ruggieri:2014bqa,Herbst:2010rf,Khan:2016exa}. 
The aforementioned chiral crossover is reproduced by the QM model, with $T_c$ 
in the same ballpark of the QCD one.
At $\mu\neq 0$  
the smooth crossover of the QM model becomes a first order phase transition, suggesting 
a critical endpoint (CEP) in the $\mu-T$ plane at which the crossover becomes a second order phase transition with divergent
susceptibilities, marking the separation between the crossover on the one hand and a first order line
on the other hand. Similar conclusions have been obtained within the Nambu-Jona-Lasinio 
model \cite{Nambu:1961tp,Nambu:1961fr,Vogl:1991qt,Klevansky:1992qe,Hatsuda:1994pi,Buballa:2003qv,Asakawa:1989bq}.
Moreover, in recent years the picture of the QM model has been strengthened by the inclusion of critical fluctuations
via the Functional Renormalization 
Group \cite{Zhang:2017icm,CamaraPereira:2020xla,Barnafoldi:2016tkd,Tripolt:2017zgc}.

Besides $\mu$, it has been recently suggested that another chemical potential,
$\mu_H$, conjugated to the helical density, $n_H$, can be of interest for the structure of QCD, see
\cite{Chernodub:2020yaf,Ambrus:2019ayb,Ambrus:2019khr,Pal:2010ih,Kapusta:2019ktm,Ambrus:2020oiw}.
Matter with $n_H\neq 0$ is called {\em helical matter} and is opposed to normal quark matter in which baryonic density
$n_B\neq 0$ and $n_H=0$.
The very basic idea of helical matter is that helicity is a conserved quantities for free massless as well as massive fermions;
therefore, if a net helicity is formed in a system made of free quarks, a chemical potential, $\mu_H$, can be introduced
that is conjugated to  $n_H$, similarly to the baryon chemical potential that is conjugated to the baryonic density.
For the quark-gluon plasma produced in high energy nuclear collisions, lumps of matter with $\langle n_H^2\rangle\neq 0$
can be formed thanks to event-by-event fluctuations, and the helical density is approximately conserved because
helicity changing processes take place on a time scale that typically is larger than the lifetime of the quark-gluon plasma 
itself \cite{Manuel:2015zpa}.
The fact that matter with helical density can be of relevance for high energy nuclear collisions
has been discussed very recently
\cite{Chernodub:2020yaf,Ambrus:2019ayb,Ambrus:2019khr,Pal:2010ih,Kapusta:2019ktm,Ambrus:2020oiw}.

We report on helical matter both at zero and finite temperature.  $\mu_H$
does not introduce additional divergences,
therefore the renormalization of the vacuum quark loop  can be done in the standard 
manner \cite{Zhang:2019neb,Frasca:2011zn,Ruggieri:2013cya,Ruggieri:2014bqa,Skokov:2010sf}.
We discuss the transition from the hadron gas to normal quark matter and helical matter.
We focus on the baryonic and the helical density susceptibility.
We show that close to the phase transition, it is possible to distinguish between normal and helical matter
not only in terms of the densities, but also in terms of the susceptibilities that indeed measure the 
fluctuations of the conserved quantities. Our results have some overlap with \cite{Chernodub:2020yaf},
and we agree with that reference whenever we compute similar quantities.
One of our conclusions is that helical density makes the chiral
phase transition softer. Thus, the formation of lumps of helical matter in the quark-gluon plasma
produced in high energy nuclear collisions can impact the observables for the critical endpoint,
since for regions in which $\langle n_H^2\rangle\neq 0$ the critical endpoint would move to
larger values of the baryon chemical potential and smaller temperatures.

We complete our study by suggesting a coupling of the fluctuations of $n_H$
and the angular velocity, $\omega$, of a rigid rotation. 
Rotation has been considered in the context of hot quark matter several times in recent literature,
see \cite{Jiang:2018wsl,Chen:2015hfc,Ebihara:2016fwa,Chernodub:2016kxh,Ambrus:2019ayb,Iyer:1982ah,
Ambrus:2015lfr,Yamamoto:2013zwa,Jiang:2016wvv,Vilenkin:1980zv,Wang:2018sur,Fetter:2009zz,Ambrus:2014uqa}
and references therein.
This is the first time in which rotation is considered within helical matter,  
therefore we limit ourselves to the simplest implementation of rotation, namely that of an unbounded system 
and we study only quark matter
in proximity of the rotation axis. We illustrate, by using an ideal gas as an example, that rotation links to the fluctuations
of the helical density and enhances them. We compute analytically the relation between the helical susceptibility and
$\omega$ in the case of a massless gas. Then we show how this link appears within the QM model.
We think that also this part of the study motivates the relevance of helical matter for the quark-gluon plasma
produced in high energy nuclear collisions, due to the vorticity produced in the collisions
\cite{Liang:2004ph,Huang:2011ru,Huang:2015oca,
Becattini:2007nd,Becattini:2013fla,
Jiang:2016woz,Aristova:2016wxe,Deng:2016gyh,Becattini:2015ska} that enhances the fluctuations of $n_H$.

The plan of the article is the following. In Section II we formulate the QM model with a helical chemical potential
and the renormalization of the vacuum term.
In Section III we discuss the phase diagram at $T=0$. In Section IV we present the results at finite temperature.
In Section V we discuss the enhancement of the fluctuations of helical density in a rotating medium.
Finally, in Section VI we summarize our conclusions and an outlook of possible future projects.
 We use the natural units system $\hbar=c=k_B=1$ throughout this article.
 
 \section{The quark-meson model for helical matter\label{sec:renPPP}}
In this section we summarize the quark-meson (QM) model that we use in our study. We firstly define
the helical chemical potential, $\mu_H$, that we couple to quarks beside the standard quark number chemical potential, $\mu$.
Then we define the lagrangian density of the model. Finally, we specify the thermodynamic grand potential and 
its renormalization procedure.

\subsection{Helical chemical potential}
We begin with a free, massive Dirac fermion.
Helicity is defined, for momentum eigenstates with eigenvalues $\bm p$, 
as the projection of spin $\bm{s}$ over the direction of momentum $\bm{p}$, that is
\begin{equation}
h=\frac{\bm s \cdot \bm p}{|\bm p|},\label{eq:helicity1}
\end{equation}
where $s_{i}=\frac{1}{2} \varepsilon_{0ijk}\Sigma^{jk}$ is the spin operator with 
$\Sigma^{\mu \nu}=\frac{i}{4} [\gamma^{\mu},\gamma^{\nu}]$.
Helicity can take the values $h=\pm1/2$ only; 
for convenience, we rescale the helicity operator by a factor of 2, as $\eta=2 h$ and 
get $\eta \psi = \aleph \psi $.  Thus, the eigenvalues  $\aleph$ are now $\pm 1$.
Helicity is relevant since it is conserved for a free Dirac fermion, and can be used beside momentum to label the energy eigenstates.
The helicity operator~\eqref{eq:helicity1} is nonlocal in coordinate space,
due to the $1/|\bm p|$: however,  a proper system of eigenfunctions can be chosen
to take the trace for the computation of the thermodynamic potential, 
namely the standard momentum/helicity eigenfunctions: with this choice
the nonlocality of $h$ is harmless and states can be labeled by the helicity eigenvalue \cite{Ambrus:2015lfr}.

The aim of our  study is to consider the effects of helicity on the QCD phase diagram:
helicity is a conserved quantity for a free massive fermion, therefore it is possible to introduce a chemical potential,
$\mu_H$, that is coupled to the net helicity of the system in equilibrium, $n_H = N_+ - N_-$ where
$N_\pm$ denotes the number of particles with $\aleph=\pm 1$. For the Dirac field the relevant operator is
\begin{equation}
\hat N_h = \bar\psi\gamma^0\eta\psi,
\end{equation}
analogous to the quark number operator,
\begin{equation}
\hat N = \bar\psi\gamma^0\psi.
\end{equation} 
It is an obvious fact that in hot quark matter, as the one produced in high energy nuclear collisions, 
the helicity of quarks can be changed by microscopic processes even though regions
with $\langle n_H^2\rangle\neq 0$ are formed by event-by-event fluctuations;
however, these processes are known to happen for time scales larger than the typical lifetime of the fireball \cite{Manuel:2015zpa},
therefore for the purpose of describing the matter produced in the collisions,
net helicity is fairly conserved and it is possible to introduce a chemical potential, $\mu_H$, 
conjugated to $\hat N_h$ in the model.  

The Lagrangian density of a free, massive Dirac fermion with both helical and number chemical potential, $\mu$, 
in momentum space is
\begin{equation}
 \mathcal{L}=\overline{\psi}( \slashed{p}+\mu_{H}\gamma^{0} \eta + \mu \gamma^{0} + m)\psi.
 \label{Eq.Lagrangianwithhelicity}
\end{equation}
The poles of the propagator in momentum space define the energy spectrum,
\begin{equation}
 p_{0, \aleph}^{(s)}(\bm{p})= \sqrt{\bm{p}^{2}+m^{2}}-s\mu -\aleph \mu_{H},
 \end{equation}
with $s=\pm1$ and $\aleph=\pm 1$.

\subsection{Lagrangian density of the quark-meson model}
The QM model consists of a mesons and quarks that interact  via an $O(4)-$invariant local interaction term.
The meson part of the Lagrangian density of the QM model is
\begin{equation}
\mathcal{L}_\textrm{mesons}=\frac{1}{2}(\partial^{\mu} \sigma \partial_{\mu} \sigma + \partial^{\mu} \bm{\pi} \partial_{\mu} \bm{\pi})- \frac{\lambda}{4}(\sigma^{2}+\bm{\pi}^2-v^2)^2-h\sigma
\label{Eq.MesonTerm}
\end{equation}
where $\bm{\pi}=(\pi_{1},\pi_{2},\pi_{3})$ corresponds to the pion isotriplet field. This Lagrangian density is invariant under $O(4)$ rotations. On the other hand, as long as $v^2>0$ the potential develops an infinite set of degenerate minima. 
Among these minima, we choose the ground state to be the one corresponding to
\begin{equation}
\langle \bm{\pi} \rangle =0, \langle \sigma \rangle \neq 0,
\label{Eq.GroundState}
\end{equation}
since $\langle \bm{\pi} \rangle \neq0$ would correspond to a spontaneous breaking of parity which is not observed
in QCD, unless an isospin chemical potential is introduced, see \cite{Mannarelli:2019hgn} for a review. 
The ground state specified in Eq.~(\ref{Eq.GroundState}) breaks the $O(4)$ symmetry down to $O(3)$ since the vacuum is invariant only under the rotations of the pion fields. The term 
\begin{equation}
\mathcal{L}_{\textrm{mass}}=h \sigma
\end{equation}
in Eq.~\eqref{Eq.MesonTerm} is responsible of the explicit breaking of chiral symmetry, where $h=F_\pi m_\pi^2$.

The quark sector of the QM model is described by the Lagrangian density
\begin{equation}
\mathcal{L}_{\textrm{quarks}}=\overline{\psi}(i\partial_{\mu} \gamma^{\mu}-g(\sigma+i\gamma_{5}\bm{\pi} \cdot \bm{\tau}))\psi
\label{Eq.QuarkTerm}
\end{equation}
where $\bm{\tau}$ are Pauli matrices in the flavour space. 
Quarks get a constituent mass because of the spontaneous breaking of the $O(4)$ symmetry in the meson sector,
\begin{equation}
M_q = g\langle\sigma\rangle.\label{eq:jjj}
\end{equation} 
In this study we limit ourselves to the mean field approximation in which $\sigma=\langle\sigma\rangle$ and
$\pi=\langle\pi\rangle$, therefore from now on anytime we write down a meson field we have in mind its expectation value
and we will skip the $\langle\rangle$ for the sake of notation. 
Finally,  the lagrangian density of the model is given by
\begin{equation}
\mathcal{L}_{\textrm{QM}}=\mathcal{L}_{\textrm{quarks}}+\mathcal{L}_{\textrm{mesons}}.
\end{equation}

\subsection{Thermodynamic potential}
The mean field effective potential of the QM model at zero temperature is given by
\begin{equation}
\Omega = U + \Omega_{q} + \Omega_T,\label{eq:OmegaTotNNN}
\end{equation}
where
\begin{equation}
U = \frac{\lambda}{4} (\sigma^2 + \bm{\pi}^2 -v^2)^2 - h \sigma
\label{Eq.EffectivePotential}
\end{equation}
is the classical potential of the meson fields as it can be read from Eq.~\eqref{Eq.MesonTerm} while $\Omega_q + \Omega_T$
corresponds to the sum of the vacuum and thermal contributions of the quarks. 

%\begin{widetext}
Firstly we analyze the vacuum term,
\begin{equation}
\Omega_{q}=-2N_{c}N_{f} \int \frac{d^{3}p}{(2\pi)^3} E_{p},
\label{Eq.QuarkLoop}
\end{equation}
with $E_{p}=\sqrt{p^2+M_q^2}$
and $M_q$ denotes the constituent quark mass defined in Eq.~\eqref{eq:jjj}.
Equation~\eqref{Eq.QuarkLoop} corresponds, after renormalization, to the quarks contribution to the condensation energy
at $T=\mu=\mu_H=0$.
Following the renormalization procedure of \cite{Zhang:2019neb} we can replace 
$\Omega_q\rightarrow\Omega_q^\mathrm{ren}$ 
in Eq.~\eqref{eq:OmegaTotNNN}, with
\begin{eqnarray}
\Omega_{q}^{\textrm{ren}} &=& \frac{3 g^4}{32\pi^2}N_{c}N_{f}\sigma^{4}
-\frac{g^4 F^{2}_{\pi}}{8\pi^{2}}N_{c}N_{f}\sigma^{2}\nonumber\\
&&+\frac{g^4 N_{c} N_{f}}{8\pi^2}\sigma^{4} \textrm{log} \frac{F_{\pi}}{\sigma}.
\label{eq:renKLM}
\end{eqnarray}
The finite temperature thermodynamic potential is standard: following \cite{Chernodub:2020yaf} to take into account
the helical as well as the number chemical potentials, we get
 \begin{eqnarray}
 \Omega_{T}&=& -\frac{ N_c N_f}{\beta} \sum_{s=\pm1}\sum_{\aleph\pm1}\nonumber\\
 &&\times\int\frac{d^3p}{(2\pi)^3} 
 \ln\left(
 1+e^{-\beta( E_p-s \mu -\aleph \mu_{H})}
 \right),
 \label{Eq.ThermodynamicPotential1}
 \end{eqnarray}
where $\beta=1/T$ and the summations run over the signs of energy and helicity. 
Note that due to the summation over $s$ and $\aleph$ in Eq.~\eqref{Eq.ThermodynamicPotential1},
the thermodynamic potential at $\mu=0$ and $\mu_H\neq 0$ is equivalent
to that at $\mu\neq 0$ and $\mu_H=0$. This duality has been observed noticed in  \cite{Chernodub:2020yaf}
for the first time.

For the sake of illustration, we write down the relation between $n_H$ and $\mu_H$, 
$\mu$ and $T$ for the case of an ideal massless gas. This can be
obtained analytically by virtue of $\Omega_T$ that can be computed exactly in this limit, namely 
\begin{eqnarray}
\frac{\Omega_T}{N_c N_f} &=&-\frac{7\pi^2 T^4}{180}-\frac{\mu^2 T^2}{6}-\frac{\mu^4}{12\pi^2}\nonumber\\
&&-\frac{\mu_H^2 T^2}{6}-\frac{\mu_H^4}{12\pi^2}-\frac{\mu_H^2\mu^2}{2\pi^2}.
\end{eqnarray}
From this relation we get
\begin{equation}
n_H =N_c N_f \frac{\mu_H^3}{3\pi^2} + N_c N_f \mu_H \left(\frac{T^2}{3}+ \frac{ \mu^2}{\pi^2}\right).\label{eq:nHmassless}
\end{equation}  
The relation between $n_H$ and $\mu_H$ is similar to that between $\mu$ and the standard baryonic density,
\begin{equation}
n_B =N_c N_f \frac{\mu ^3}{3\pi^2} + N_c N_f \mu  \left(\frac{T^2}{3}+ \frac{ \mu_H^2}{\pi^2}\right).\label{eq:nHmasslessAAA}
\end{equation}
The duality $\mu\leftrightarrow\mu_H$ is evident by comparing Eqs.~\eqref{eq:nHmassless} and~\eqref{eq:nHmasslessAAA}.

\section{Results for helical matter at zero temperature}

Our parameter set is $M_{\sigma}=700$ MeV, $v=F_{\pi}=93$ MeV, $M_{\pi}=138$ MeV and $g=3.6$,
$\lambda=M_{\sigma}^{2}/2F_{\pi}^{2}=28.3$ and $h=M_{\pi}^{2} F_{\pi}=1.78\times10^{6}$ MeV$^{3}$; 
these give $M_q=335$ MeV in the vacuum. 
The thermodynamic potential is invariant under the changes $\mu\rightarrow -\mu$ or $\mu_{H}\rightarrow -\mu_{H}$, therefore
the signs of two chemical potentials are irrelevant and we consider $\mu>0$ and $\mu_{H}>0$ only.
Moreover, we have verified that changing $M_\sigma$ down to $M_\sigma=550$ MeV does not alter the results qualitatively,
therefore we limit ourselves to report only the results obtained for $M_\sigma=700$ MeV.

\subsection{The phase diagram at zero temperature}

\begin{figure}[t!]
\begin{center}
\includegraphics[scale=0.3]{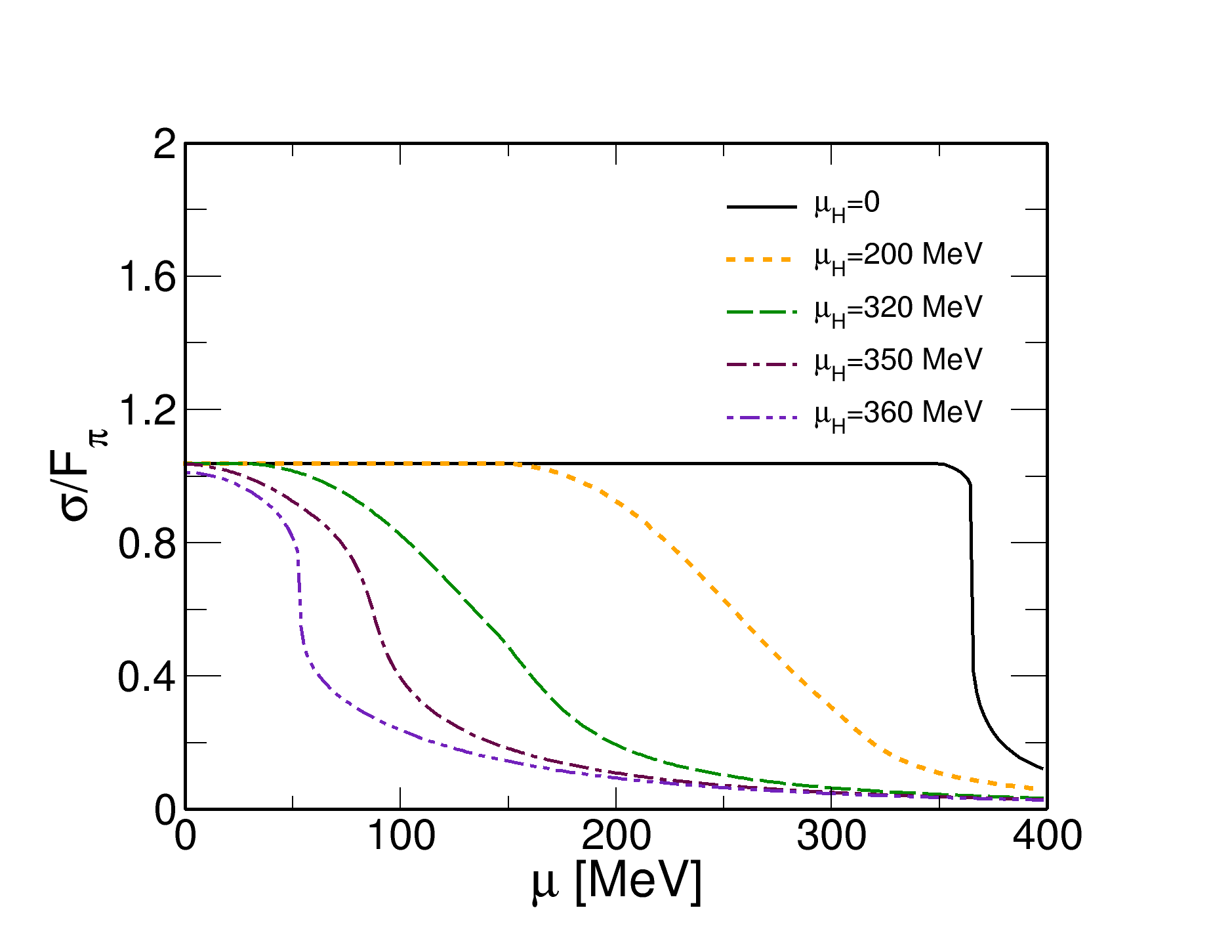}\\
\includegraphics[scale=0.3]{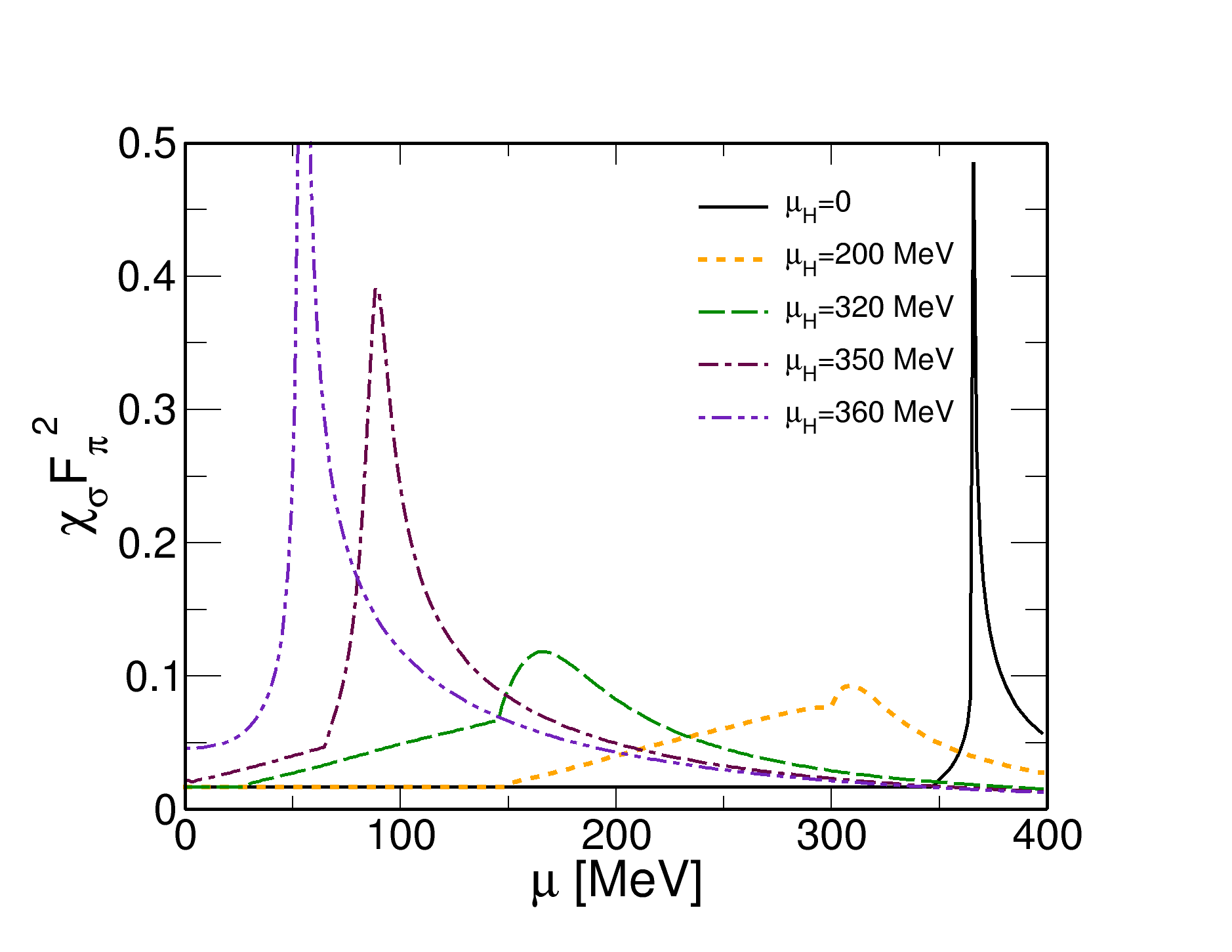}
\end{center}
\caption{\label{Fig:ams}Condensate (upper panel) and chiral susceptibility (lower panel) versus $\mu$ at $T=0$
and for several values of $\mu_H$.}
\end{figure}

The phase diagram of a system with unbalanced helicity at $T=0$ has been studied in~\cite{Chernodub:2020yaf} for the first time:
here we confirm the results of~\cite{Chernodub:2020yaf} and in addition to this, we analyze how the vacuum renormalized term of the
thermodynamic potential affects the picture.
In the upper panel of Fig.~\ref{Fig:ams} we plot the condensate versus $\mu$ at $T=0$ and for several values of $\mu_H$.
At $\mu_H=0$ a first order phase transition for $\mu=\mu_c\approx 365$ MeV 
from the chiral broken phase to the chiral restored phase has been found
in agreement with previous model calculations~\cite{Zhang:2019neb}: the chiral restored phase can be dubbed as normal quark 
matter phase since quark number density is finite while helical density is vanishing.
Increasing $\mu_H$ we notice that the phase transition becomes smoother and eventually the discontinuity
of the order parameter disappears: a smooth crossover replaces the first order phase transition in some range of $\mu_H$.
In addition to this, the critical value of $\mu$ becomes smaller. 
Increasing $\mu_H$ even further the crossover becomes stronger again and turns to another first order phase transition;
eventually, at $\mu=0$ and $\mu_H=\mu_c$ a first order transition happens 
to a new phase in which chiral symmetry is restored
but quark matter has vanishing density and finite helical density; the fact that $\mu_c$ sets the scale for the phase transition
in the two cases is due to the duality $\mu\leftrightarrow\mu_H$ of the thermodynamic potential.

The smoothening followed by the hardening of the phase transitions when $\mu_H$ is increased
is evident when we compute the chiral susceptibility, \cite{Fukushima:2003fw,Sasaki:2006ww,Sasaki:2006ws,Abuki:2008nm}, 
\begin{equation}
\chi_\sigma = \left(\frac{\partial^2\Omega}{\partial\sigma^2}\right)^{-1}
\end{equation}
around the transitions. The results are shown in the lower panel of Fig.~\ref{Fig:ams}
in which we plot the dimensionless quantity $F_\pi^2 \chi_\sigma$ versus $\mu$ for several values of $\mu_H$.
We find that for $\mu\approx\mu_c$ or $\mu_H \approx \mu_c$ the susceptibility is very peaked around the transition,
while the peaks become smaller and broader as we move towards the region $\mu\approx \mu_H$,
signaling that the transition turns to a smooth crossover.

\begin{figure}[t!]
\begin{center}
\includegraphics[scale=0.3]{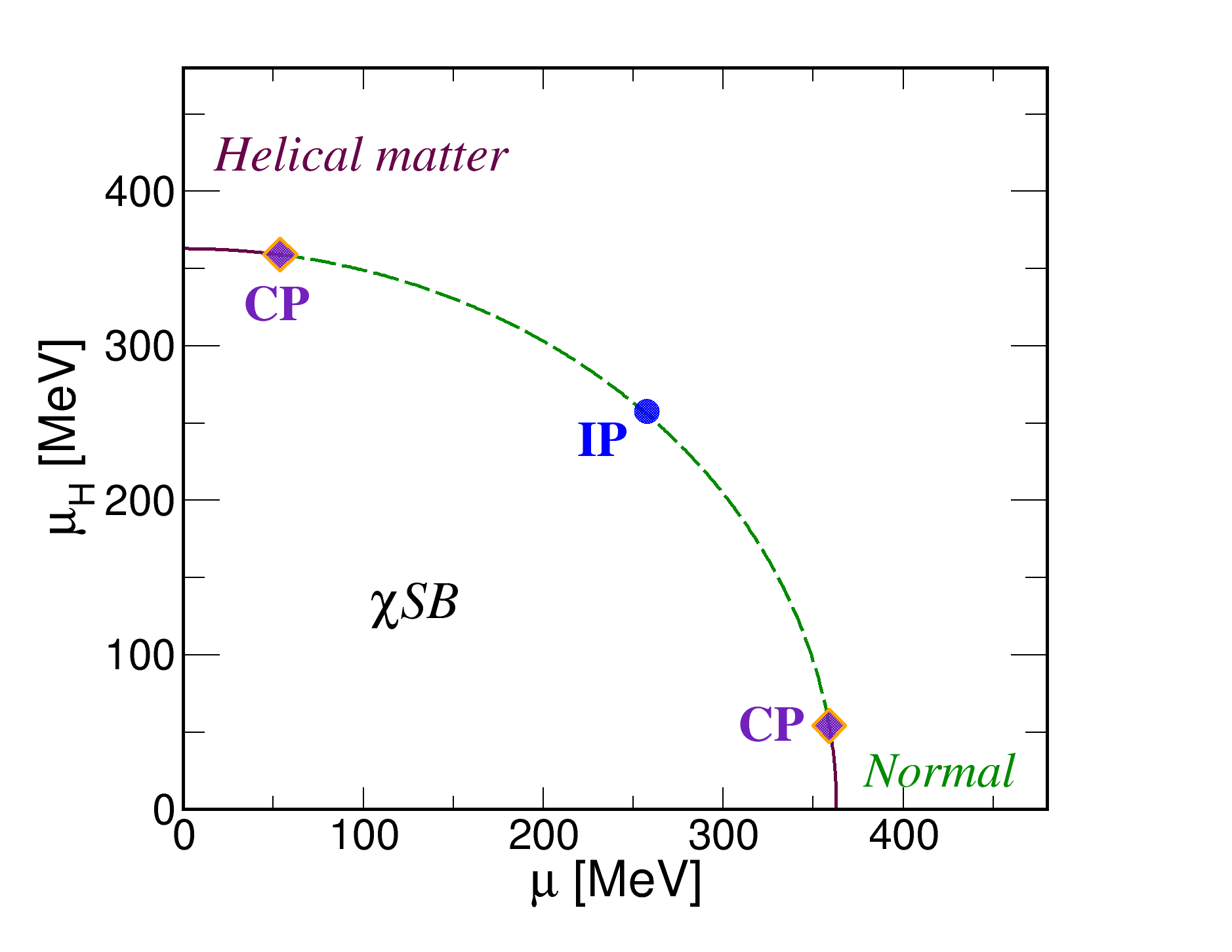}
\end{center}
\caption{\label{Fig:pdt0}Phase diagram in the $\mu-\mu_H$ plane at $T=0$. 
Solid lines denote first order transitions where a discontinuity of the condensate takes place,
dashed line stands for the crossover around which the condensate changes smoothly;
diamonds denote the critical endpoints (CP)
at which the crossover becomes a first order transition. The blue circle labeled IP stands for the inversion point,
namely the point at which the crossover is the smoothest realized in the phase diagram,
corresponding to the values of $\mu=\mu_H$ at which moving 
upwards or downwards along the transition line, the crossover becomes harder. 
The $\chi$SB window denotes the region where chiral symmetry is broken spontaneously;
the other regions correspond to phases with finite quark number density, $n$, and/or 
helical density, $n_H$.
In the {\it normal} matter chiral symmetry is restored and $n\gg n_H$,
while {\it helical matter} denotes the region where chiral symmetry is restored but $n_H\gg n$.}
\end{figure}

In Fig.~\ref{Fig:pdt0} we plot the phase diagram in the $\mu-\mu_H$ plane at $T=0$. 
The solid lines denote first order transitions where a discontinuity of the condensate takes place,
while the dashed line stands for the crossover around which the condensate changes smoothly.
The diamonds denote the critical endpoints
at which the crossover becomes a first order transition. The circle is the inversion point,
namely the point at which the crossover is the smoothest realized in the phase diagram,
corresponding to the values of $\mu=\mu_H$ at which moving 
upwards or downwards along the transition line, the crossover becomes harder.
All the lines have been obtained from the results shown in 
Fig.~\ref{Fig:ams} by identifying the critical values of $\mu$ with those at which the chiral susceptibility
has the peak. Clearly the transition line is symmetric under the exchange $\mu \leftrightarrow \mu_H$.
We notice the presence of two critical points in the phase diagram, resulting from the aforementioned duality of the 
thermodynamic potential. The inversion point, IP,  at which the crossover is the smoothest is $\mu = \mu_H \approx 257$ MeV.

The chiral symmetry restored phases in the lower right and upper left corners of
the phase diagrams are quite different: for $\mu\gg\mu_H$ quark matter has 
both number and helical densities, denoted by $n$ and $n_H$ respectively, but $n\gg n_H$
therefore it is legit to call this phase as normal quark matter;
on the other corner of the phase diagram where $\mu_H \gg \mu$ the 
dual situation $n_H \gg n$ is realized, therefore this phase can be dubbed as helical quark matter.

\begin{figure}[t!]
\begin{center}
\includegraphics[scale=0.3]{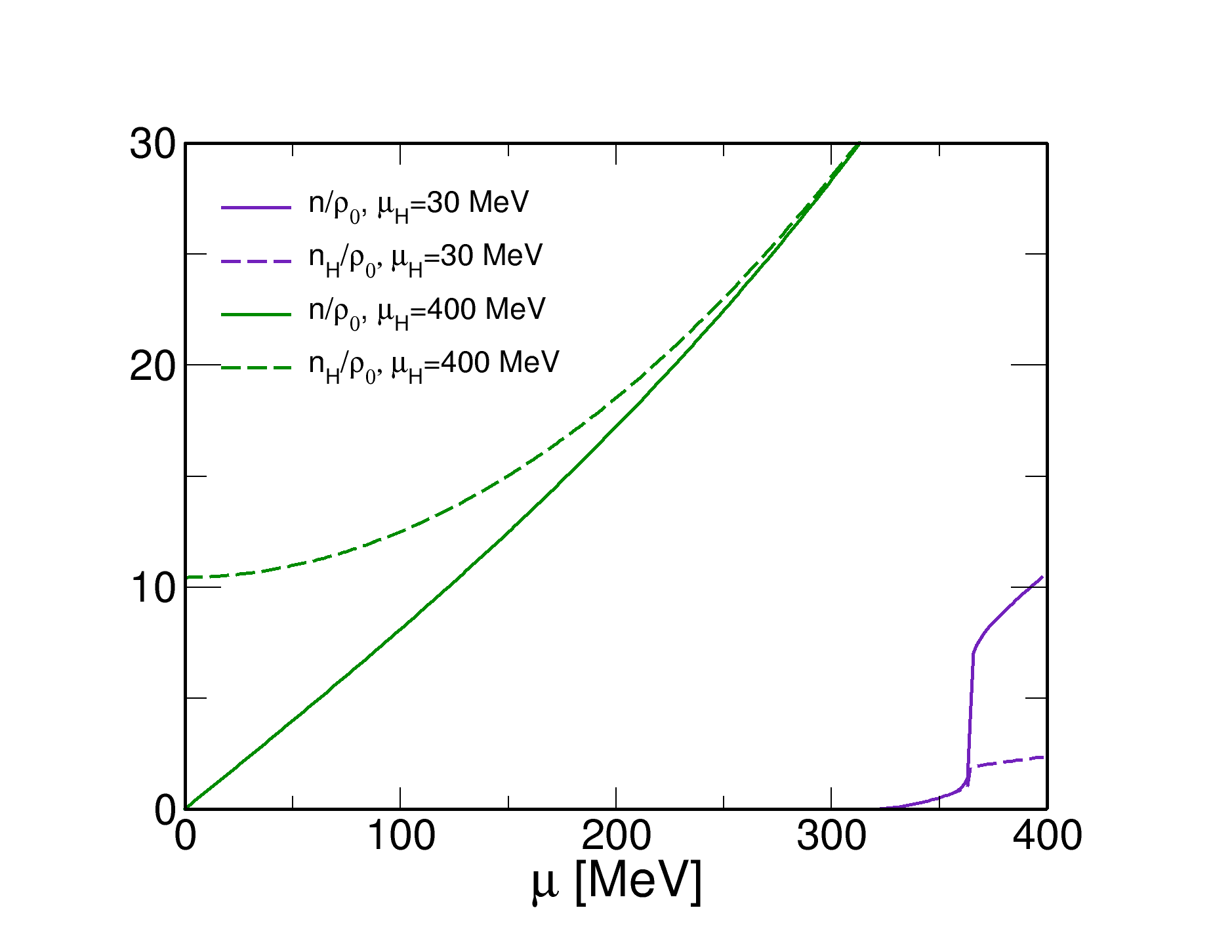}
\end{center}
\caption{\label{Fig:amsDD}Densities, measured in units of the nuclear saturation density $\rho_0=0.16$ fm$^{-3}$,
versus $\mu$. }
\end{figure}

One way to distinguish the Helical Matter from the normal phase is the value of the densities,
 \begin{eqnarray}
n &=& -\frac{\partial\Omega}{\partial\mu}, \\
n_H &=& -\frac{\partial\Omega}{\partial\mu_H},
 \end{eqnarray}
where $n$ and $n_H$ denote the quark number density and the helical density respectively. In Fig.~\ref{Fig:amsDD} 
we plot $n$ and $n_H$ versus the quark chemical potential for two representative values of $\mu_H$:
indigo lines correspond to $\mu_H=30$ MeV to cover the lower right corner of the phase diagram
in Fig.~\ref{Fig:pdt0}, and green lines to $\mu_H=400$ MeV that cover the upper left corner of the phase diagram
and extend then to higher values of $\mu$. The normal phase is characterized by $n \gg n_H$ while
the Helical Matter has $n_H \gg n$. Increasing $\mu$ in the case of large $\mu_H$ the system smoothly connects
to a phase in which $n \approx n_H$.
 
\begin{figure}[t!]
\begin{center}
\includegraphics[scale=0.3]{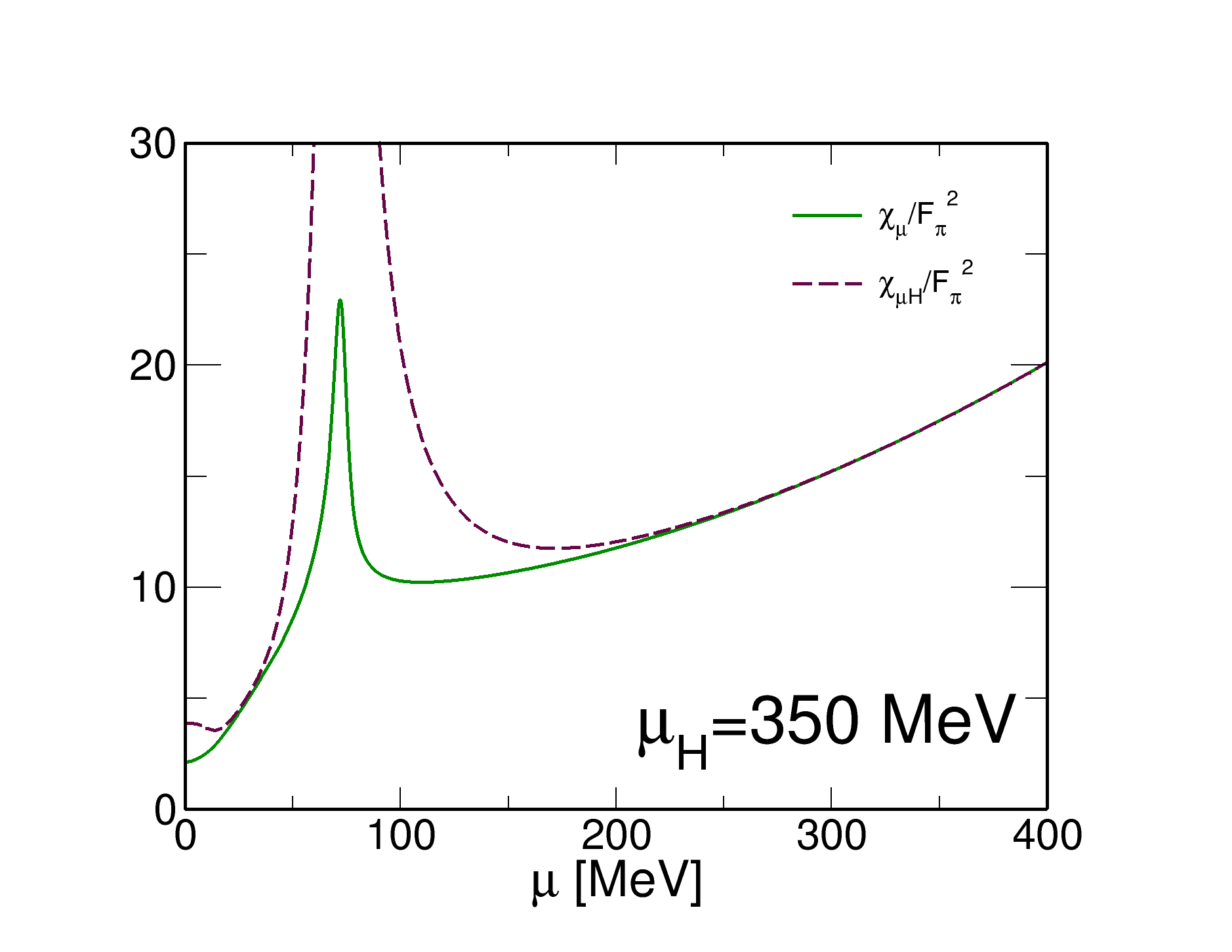}\\
\includegraphics[scale=0.3]{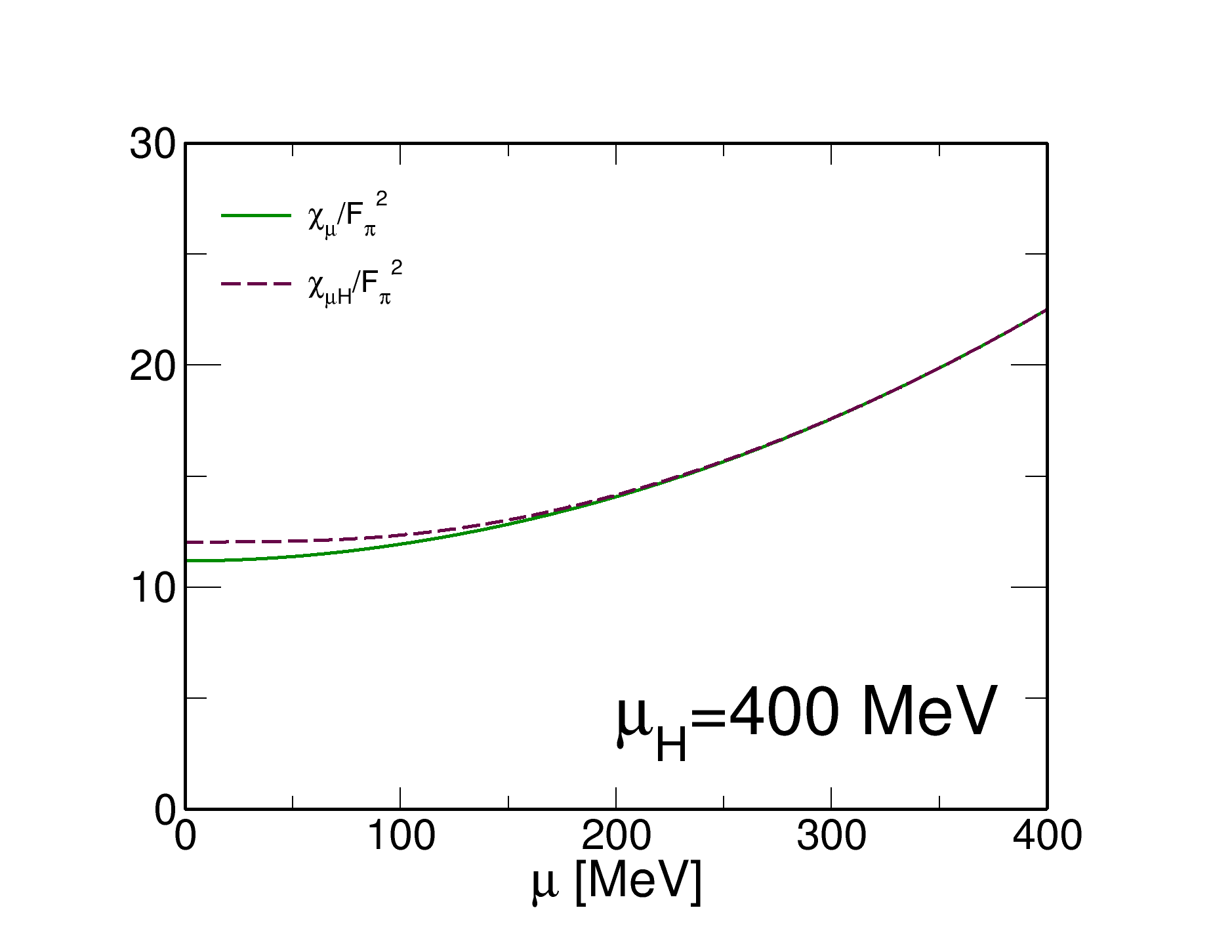}
\end{center}
\caption{\label{Fig:amsHH}Condensate (upper panel) and chiral susceptibility (lower panel) versus $\mu$ at $T=0$
and for several values of $\mu_H$.}
\end{figure}

In Fig.~\ref{Fig:amsHH} we plot the susceptibilities of quark number density, $\chi_\mu$,
and of helical density, $\chi_{\mu H}$, defined respectively as
\begin{eqnarray}
\chi_\mu &=& - \frac{\partial^2\Omega}{\partial\mu^2},\label{eq:dabra}\\
\chi_{\mu H} &=& - \frac{\partial^2\Omega}{\partial\mu_H^2}.\label{eq:abra}
\end{eqnarray}
In the upper panel of Fig.~\ref{Fig:amsHH}  we have collected the results for $\mu_H=350$ MeV,
while in the lower panel $\mu_H=400$ MeV: the two values of
$\mu_H$ that allow us to probe the Helical Matter phase crossing the transition line and above the transition line respectively.
We notice that around the crossover line, $\chi_{\mu H}$ has a clear peak structure while the peak of $\chi_\mu$ is very modest,
indicating that in that region the fluctuations of the helical density are much more pronounced that those of the
quark number. On the other hand, increasing $\mu$ and getting far from the crossover line, the two susceptibilities
agree with each other: quark number and helicity fluctuate in the same way.
This is clearer for $\mu_H=400$ MeV in which case no transition line is crossed: the two susceptibilities
agree with each other in the full range of $\mu$ studied, denoting that the fluctuations of
the quark number and of the helical density are the same.
We have checked that the dual situation happens moving at low $\mu_H$ and crossing the critical lines increasing $\mu$.

We have verified that changing the value of $M_\sigma$ in the range $(550,700)$ MeV 
does not lead to significant changes in the condensates and in the phase diagram.
Moreover, lowering the value of $g$ the phase transitions tend to be smoother
and eventually for $M_q=g F_\pi=300$ MeV at $T=\mu=\mu_H=0$, the first order lines disappear.

\begin{figure}[t!]
\begin{center}
\includegraphics[scale=0.3]{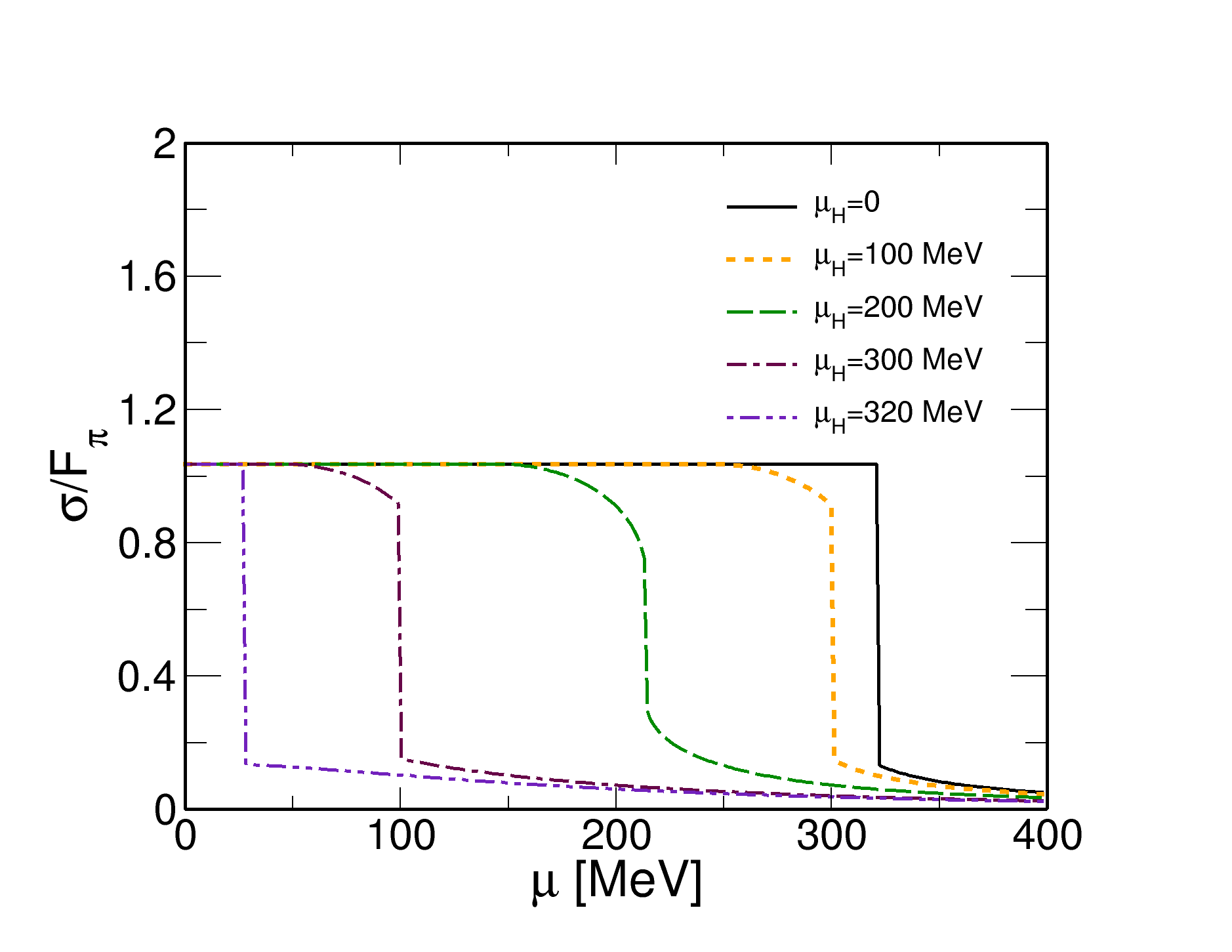}\end{center}
\caption{\label{Fig:amsW}Condensate  versus $\mu$ at $T=0$
and for several values of $\mu_H$, computed for the model without the quark renormalized vacuum term in the thermodynamic potential.}
\end{figure}

We have analyzed the effect of removing the renormalized vacuum term of quarks
in the thermodynamic potential, see Eqs.~\eqref{eq:OmegaTotNNN} and~\eqref{eq:renKLM}.
In Fig.~\ref{Fig:amsW} we plot the condensate versus $\mu$ at $T=0$ for several values of $\mu_H$,
for the model without the vacuum term.
Firstly, we notice that at $\mu_H=0$ the chiral phase transition happens for $\mu= \mu_c \approx 320$ MeV,
which is smaller than the critical chemical potential that we have found when we have included the vacuum term,
see Fig.~\ref{Fig:ams}: the vacuum term makes the chiral symmetry broken phase more stable.
Then, we notice that the phase transition remains of the first order
for any value of $\mu_H$;
we notice that changing $\mu_H$ from zero to
about $200$ MeV, the jump of the order parameter decreases, then it increases for larger $\mu_H$.
We interpret this as a softening of the phase transition for small $\mu_H$, which then becomes harder
as an inversion point is crossed, similarly to the one in Fig.~\ref{Fig:pdt0}.
However, we don't find any range of $\mu_H$ in which the transition turns to a smooth crossover:
the phase diagram in this case will differ from the one shown in Fig.~\ref{Fig:pdt0}
because in the former case there is only a first order phase transition line.
We conclude that the existence of the critical endpoints in the zero temperature phase diagram depends
on how the regularization of the divergent term of the thermodynamic potential is done;
however, the softening followed by the hardening of the phase transition is unaffected by this choice.

\section{Results for helical matter at finite temperature} 

\subsection{Chiral symmetry breaking}
\begin{figure}[t!]
  \centering
  \includegraphics[width=0.47\textwidth]{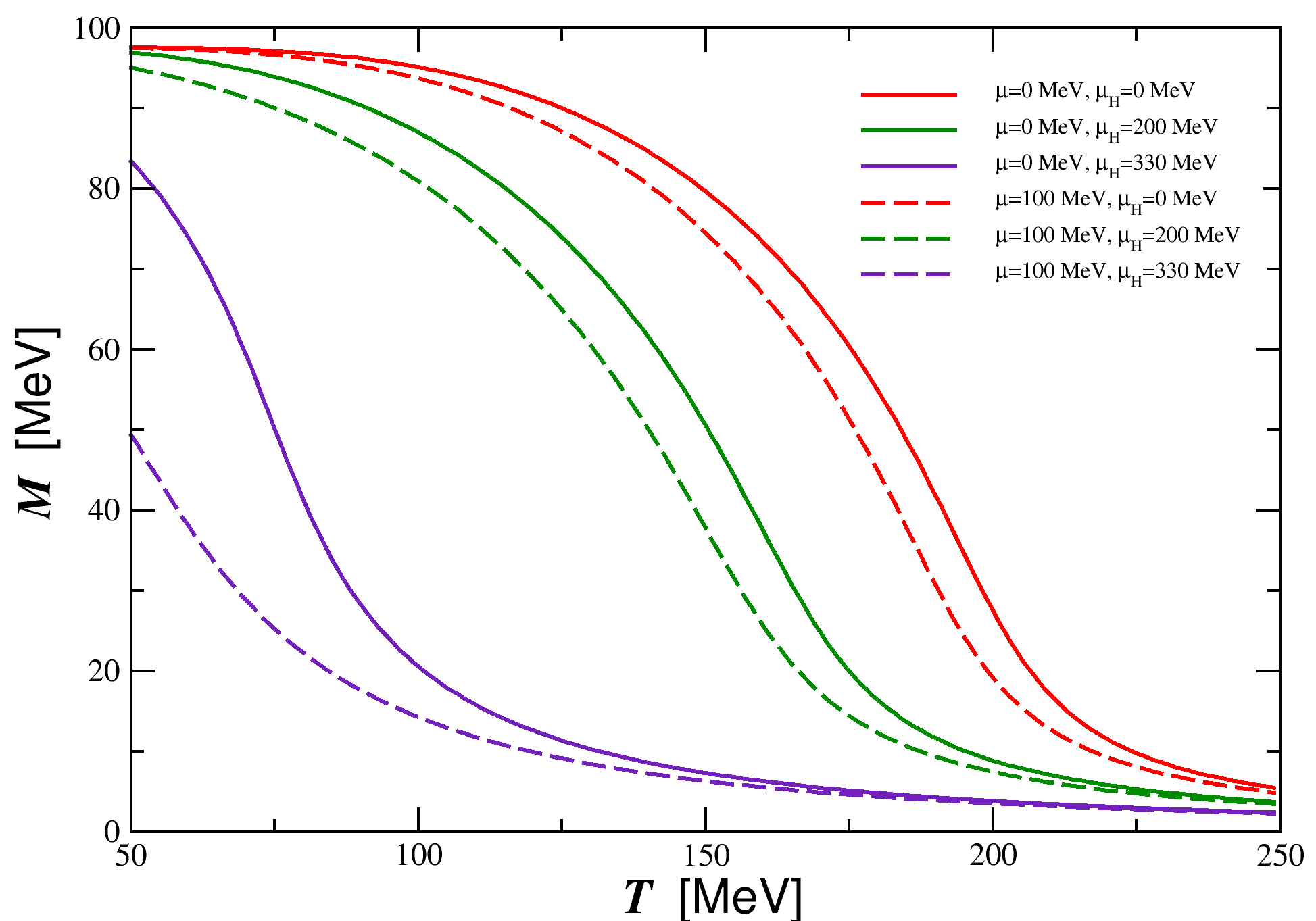}\\
  \caption{Constituent quark mass, $M$, versus temperature for several ($\mu,\mu_{H}$) pairs.
  Solid and dashed lines correspond to $\mu=0$ and $\mu=100$ MeV respectively.}
  \label{Fig.masscomparison}
\end{figure}

In Fig.~\ref{Fig.masscomparison} we plot the $\sigma-$condensate versus temperature, for several values
of $\mu$ and $\mu_H$. In general, there is a temperature range in which the condensate drops down,
signaling the approximate restoration of chiral symmetry via a smooth crossover; the pseudocritical temperature
can be identified with the temperature at which $d\sigma/d\beta$ is maximum. 
Increasing $\mu_H$ at $\mu=0$ results in the lowering of the critical temperature. This agrees with
the expectations from the zero temperature phase diagram in Fig.~\ref{Fig:pdt0},
since it is shown there that increasing $\mu_H$ at $\mu=0$ makes the chiral broken phase less stable
and eventually leads to chiral symmetry restoration and helical matter.
Therefore, it is expected that increasing $\mu_H$ at finite temperature will also lower the $\sigma-$condensate
and chiral symmetry restoration will be facilitated.   
Increasing $\mu$ shifts the pseudocritical temperature to lower values: once again, this is in agreement
with the expectation from the zero temperature phase diagram in Fig.~\ref{Fig:pdt0},
because increasing $\mu$ has the effect to destabilize the chiral symmetry broken phase.

\subsection{Phase diagram}
\begin{figure}[t!]
  \centering
  \includegraphics[width=0.5\textwidth]{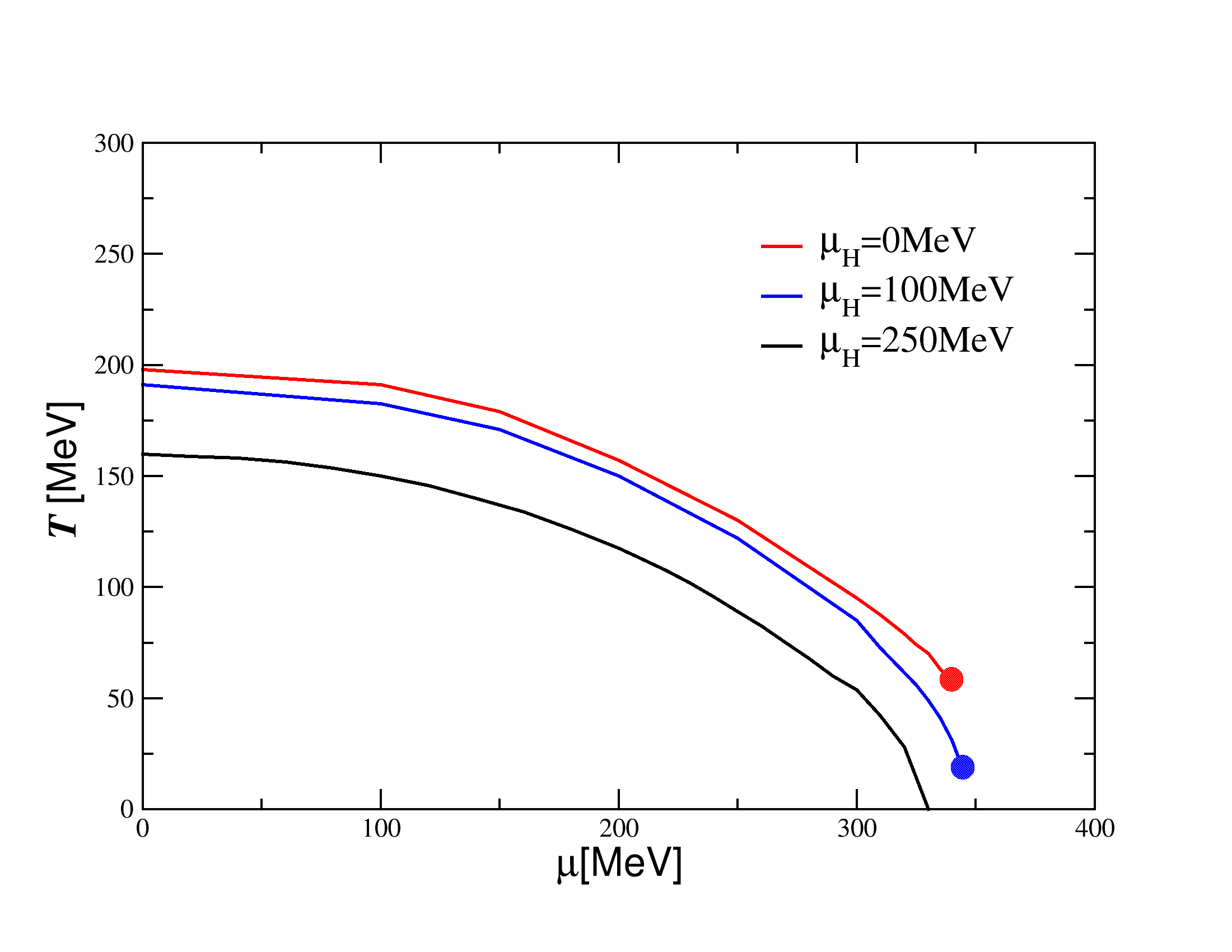}
  \caption{Pseudo-critical lines in the $\mu-T$ plane, for several values of $\mu_H$. The dot denotes the critical endpoint.}
  \label{Fig:PDpd}
\end{figure}

These results can be summarized in a phase diagram in the $\mu-T$ plane, see Fig.~\ref{Fig:PDpd}.
The solid lines denote the smooth crossover froom the chiral broken phase to the chiral symmetric phase,
computed at several values of $\mu_H$.
Firstly, increasing $\mu_H$ moves the critical line downwards, in agreement with the previous discussions.
The dots in the figures denote the critical endpoint, CEP, namely the values of $\mu$ and $T$ at which the crossover
becomes a second order phase transition with divergent susceptibilities; increasing $\mu$ further
the critical line becomes a first order, with a discontinuity of the $\sigma-$condensate.
One of the effects of increasing $\mu_H$ is to move the CEP downwards; eventually, the CEP disappears from the
phase diagram. This agrees with Fig.~\ref{Fig:pdt0} since at $T=0$ there is a large window in the $\mu-\mu_H$ plane
in which the transition to helical/normal quark matter happens via a smooth crossover rather than a first order phase transition.

\subsection{Quark number susceptibility}

\begin{figure}[t!]
  \centering
  \includegraphics[width=0.5\textwidth]{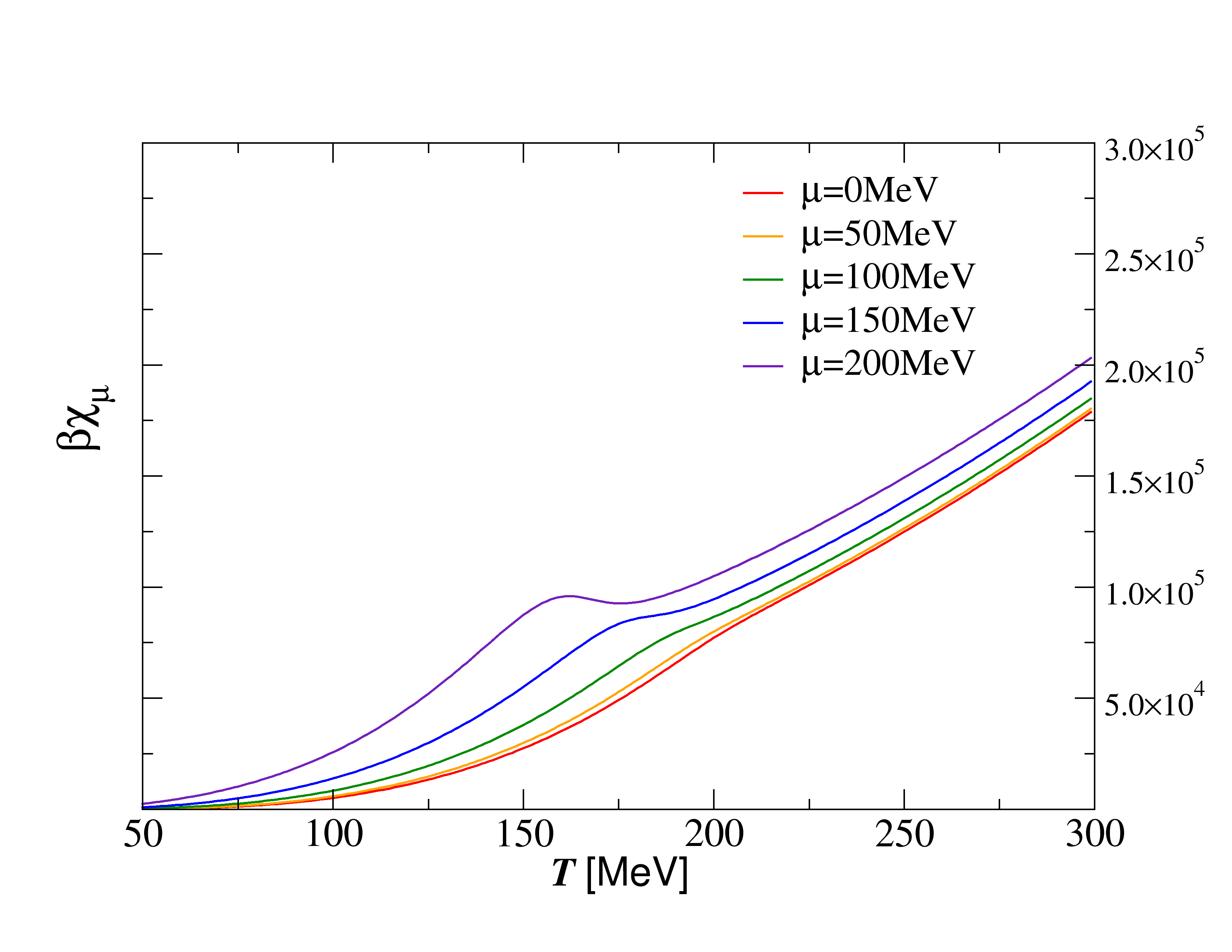}\\
    \includegraphics[width=0.5\textwidth]{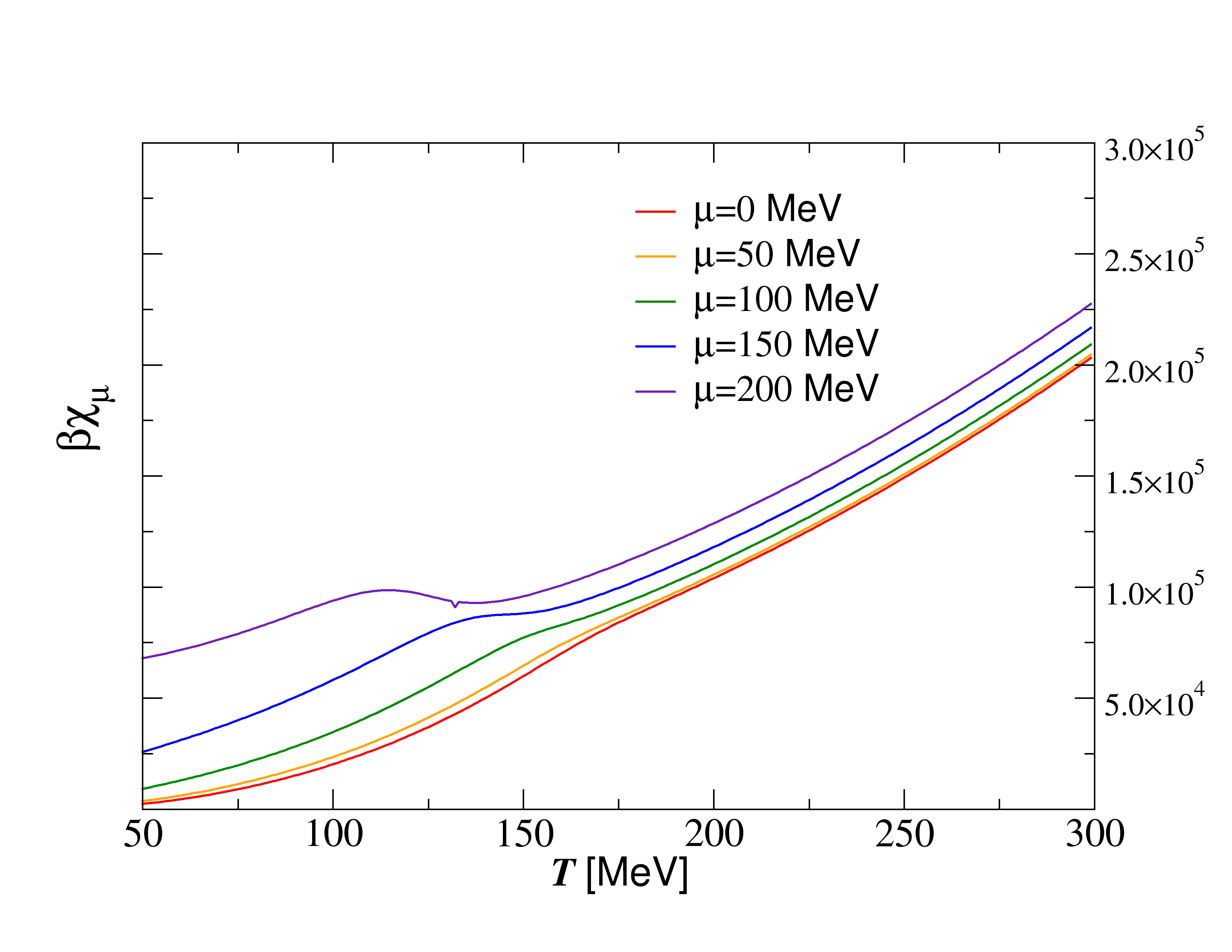}
  \caption{$\beta\chi_\mu$ (in MeV) over temperature, versus temperature, for several values of $\mu$.
  Upper and lower panel correspond to $\mu_H=0$ and $\mu_H=200$ MeV respectively.}
  \label{Fig:bfg}
\end{figure}

In Fig.~\ref{Fig:bfg} we plot the particle number susceptibility, see Eq.~\eqref{eq:dabra}, normalized to temperature, 
versus temperature, for several values
of $\mu$ and $\mu_H$. 
We notice that around the chiral crossover, increasing the value of $\mu_H$ results
in a broader susceptibility: this becomes evident if we compare the cases  $\mu_H=200$ MeV  
and $\mu_H=0$, see the upper and lower panel in Fig.~\ref{Fig:bfg}.

The  results discussed in this section show that helicity softens the transition from the quark-gluon plasma to the hadron phase.
In fact, the helical chemical potential makes the chiral phase transition softer, and if $\mu_H$ is large enough the
critical endpoint disappears from the phase diagram and the transition to the chiral symmetric phase
remains a smooth crossover in the whole $\mu-T$ plane.
This picture is confirmed by the   quark number 
fluctuations
that  are smoothed by the helical chemical potential.

\section{Helical matter under rotation}
We close this study by analyzing briefly the 
correlation between rotation and the fluctuations of the helical density.
This investigation is motivated by the potential applications to relativistic heavy ion collisions.
As a matter of fact, even if the quark-gluon plasma in collisions was produced with a net $n_H=0$,
local fluctuations of $n_H$ are possible: the amount of fluctuations is measured by the susceptibility
$\chi_{\mu H}$ defined in Eq.~\eqref{eq:abra},
where the derivative takes into account the dependence of the condensate of $\mu_H$.
We want to prove that in a rotating medium the fluctuations of $n_H$ are enhanced:
this in turn  would imply that in a medium in local equilibrium, local lumps with $n_H\neq 0$ can form
and this formation is enhanced if the system is rotating.
We limit ourselves to the simplest implementation of rotation of a fermion system,
namely we assume a rigid rotation of an unbounded system with a constant and homogeneous angular velocity, $\bm\omega$,
directed along the $z-$axis.
While the choice of more realistic boundary conditions is 
feasible \cite{Ebihara:2016fwa,Chernodub:2016kxh,Ambrus:2019ayb,Ambrus:2015lfr}, these would complicate the
calculations but probably would not change qualitatively the results: therefore, 
in this first study we prefer to adopt the simplest implementation possible to highlight the effect,
leaving a complete study to a future article.

The implementation of rotation in quark matter that we adopt has been already discussed several times
in the literature, therefore we quote the results leaving the details to the original 
references \cite{Jiang:2018wsl,Chen:2015hfc,Ebihara:2016fwa,Chernodub:2016kxh,Ambrus:2019ayb,Iyer:1982ah,
Ambrus:2015lfr,Yamamoto:2013zwa,Jiang:2016wvv,Vilenkin:1980zv,Wang:2018sur,Fetter:2009zz,Ambrus:2014uqa}.
In previous works  
a coordinate-dependent condensate has been considered, within the local density approximation
in which it is assumed that local equilibrium is realized at any point in space
and that the thermodynamic potential per unit volume is given by that of a homogeneous system at any point.
Within this approximation, the gap equation for the condensate is solved at each point in space, 
then the thermodynamic potential is obtained via an integration over the whole volume.
Leaving the solution of the more complete problem to a future study, here we limit ourselves
to consider a homogeneous condensate: this approximation is not very bad,
considering that even when the coordinate dependence is introduced, 
on average the condensate varies at most of the 10$\%$ when the radial coordinate, $r$, changes from the origin up to 
$\omega r \lesssim 0.8$.
We further limit ourselves to $\mu=0$.

Within the local density approximation, the thermodynamic potential for an unbounded system
rotating with angular velocity $\bm\omega$ along the $z-$axis, temperature $T=1/\beta$ and
helical chemical potential $\mu_H$ can be obtained from \cite{Ambrus:2015lfr,Jiang:2016wvv,Wang:2018sur,Chernodub:2020yaf}, that is
\begin{eqnarray}
\Omega_{M}
		&=&-\frac{N_c N_f}{4 \beta}\int d^3 \bm{r} \sum_{n=-\infty}^\infty 
		\int \frac{dp}{8 \pi^2} ~{\cal J}_n(p_T r)
		\nonumber\\
	&\times &\sum_{s=\pm 1}\sum_{\xi=\pm 1}\sum_{\eta=\pm 1}
	 \log\left(1+e^{(s E_p+\xi(n+1/2)\omega +\eta \mu_{H})\beta}\right),\nonumber\\&& \label{eq:alignREM}
\end{eqnarray}
where we have put
\begin{equation}
\int dp =\int dp^2_T \int dp_z,
\end{equation}
with $p_{T}$ denoting the transverse momentum while $p_{z}$ is the momentum along 
the $z-$axis; moreover,
$E_{p}=\sqrt{p_z^2+p_T^2+(g \sigma)^2}$. We have also defined 
\begin{equation}
{\cal J}_n(p_T r) = J_{n}^2(p_T r)+J_{n+1}^2(p_T r),
\end{equation}
where $J_{n}(p_T r)$ denotes the first kind Bessel function of order $n$. 
Finally, $\bm{r}$ is the location from the center of the rotation. 
 
In the form of Eq.~\eqref{eq:alignREM}, $\Omega_M$  contains a  vacuum part
that depends of $\omega$. 
At difference with previous works \cite{Jiang:2016wvv,Wang:2018sur} we subtract this $\omega-$dependent contribution 
and replace it with our previously renormalized vacuum term,
see section~\ref{sec:renPPP}.  This procedure is in line with \cite{Chernodub:2016kxh,Ambrus:2015lfr},
and is justified by the fact that the vacuum term
should depend of $\omega$: an easy way to see this is that in order to experience
the rotation, at least one particle should be put on the top of the vacuum so that this particle
can feel inertial forces that appear in a rotating system; but the quantum state of vacuum plus one particle
is different from the vacuum itself. Therefore, it is not possible to have a vacuum that feels the rotation.

After the subtraction and dividing by the volume of the system, $V$,
we are left with the matter part of the thermodynamic potential per unit volume, $\Omega_T=\Omega_M/V$, that 
replaces Eq.~\eqref{Eq.ThermodynamicPotential1} and reads
\begin{eqnarray}
\Omega_{T}
		&=&- \frac{N_c N_f}{ 2\beta V}\int d^3 \bm{r} \sum_{n=-\infty}^\infty 
		\int \frac{dp}{8 \pi^2} ~{\cal J}_n(p_T r)
		\nonumber\\
	&\times & \sum_{\xi=\pm 1}\sum_{\eta=\pm 1}
	 \log\left(1+e^{-\beta (E_p+\xi(n+1/2)\omega +\eta \mu_{H})}\right).\nonumber\\&& \label{eq:subracted444}
\end{eqnarray}
It is an easy exercise to prove that for $\omega= 0$ Eq.~\eqref{eq:subracted444} is consistent with 
Eq.~\eqref{Eq.ThermodynamicPotential1}. In fact, 
$\int d^3p/(2\pi)^3 = \int dp/(8\pi^2)$ for any function that does not depend of the azimuthal angle, and
after putting $\omega=0$ in the right hand side of Eq.~\eqref{eq:subracted444}
the $n-$dependence of the integrand is confined to the ${\cal J}_n(p_T r)$ term: the summation
over $n$ can be performed by using $J_{-n}=(-1)^n J_n$ as well as the well known identity
\begin{equation}
J_0^2(x) + 2\sum_{n=1}^\infty J_n^2(x)=1,
\end{equation}
which brings an overall factor of $2$ in Eq.~\eqref{eq:subracted444}.

Under the assumption that the condensate does not depend of $\bm r$, the integration over volume
can be done exactly: considering that ${\cal J}_{n}={\cal J}_{|n|-1}$ for $n<0$ we can limit ourselves
to evaluate ${\cal J}_{n}$ for $n>0$. We take a cylinder with radius $R$ and height $h$ that develops along the
direction of $\bm\omega$: we have $V=\pi R^2 h$ and 
\begin{eqnarray}
 \int d^3\bm r {\cal J}_n(p_T r) &=&2V\left[ J_n^2(p_T R) +  J_{n+1}^2(p_T R)\right.\nonumber\\
&&\left.
-\frac{(2n+1) }{p_T R}J_n(p_T R)J_{n+1}(p_T R)\right],n\geq 0.\nonumber\\
&&\label{eq:alpja}
\end{eqnarray}
We limit ourselves to describe a region that is very close to the center of the rotation,
putting $p_T R=0$ in the argument of ${\cal J}_n$ in Eq.~\eqref{eq:subracted444}.
For $x\approx 0$ the Bessel functions behave as 
$J_0^2(x)\approx 1$, $J_n^2(x)\approx x^{2|n|}$ for $n\neq 0$, in this approximation we have to 
consider only the contribution from $J_0$ thus take into account  ${\cal J}_0={\cal J}_{-1}\approx 1$
in the sum over $n$ in Eq.~\eqref{eq:subracted444}: 
the integration in Eq.~\eqref{eq:alpja} gives $V$ for $n=-1,0$
 and the approximate $\Omega_T$ is
\begin{eqnarray}
\Omega_{T}
		&=&- \frac{N_c N_f}{ \beta } 
		\int \frac{dp}{8 \pi^2}  
		\nonumber\\
	&\times & \sum_{\kappa=\pm 1/2}\sum_{\eta=\pm 1}
	 \log\left(1+e^{-\beta (E_p+\kappa \omega +\eta \mu_{H})}\right). \label{eq:subracted444aaa}
\end{eqnarray}
In the following we use Eq.~\eqref{eq:subracted444aaa} to estimate the effect of $\omega\neq 0$
on the fluctuations of the helical density: firstly, we illustrate the idea with an ideal gas of massive particles;
then, we compute $\chi_H$ for the QM model.

\subsection{The case of an ideal gas} 
It is instructive to evaluate the effect of  $\omega$ on $\chi_H$ in the case of an ideal gas with massive particles
with mass $m$.
In this case the thermodynamic potential is given by Eq.~\eqref{eq:subracted444} only. 

In order to mimic the conditions of high energy nuclear collisions we consider  $\mu_H\approx0$.
Moreover, to make the coupling between $\omega$ and $\mu_H$ more transparent,
we consider the lowest nontrivial order in $\omega$ in the expansion of $\Omega_T$ around $\mu_H=0$ and $\omega=0$.
Considering that $\Omega_T$ is an even function of $\mu_H$ and $\omega$ we are left with
\begin{eqnarray}
\Omega_T &=& \Omega_T(\mu_H=0,\omega=0) \nonumber\\
&&+\frac{c_{2,0}}{2} \mu_H^2 +\frac{c_{0,2}}{2} \omega^2 +\frac{c_{4,0}}{4!} \mu_H^4 +\frac{c_{0,4}}{4!} \omega^4 \nonumber\\ 
&& +\frac{c_{2,2}}{4}\omega^2\mu_H^2 \nonumber\\
&&+O(\mu_H^6,\omega^6), \label{eq:njk}
\end{eqnarray}
where   
\begin{eqnarray}
c_{m,n} &=& \left.\frac{\partial^{(m+n)}\Omega_T}{\partial\mu_H^m\partial\omega^n}\right|_{\mu_H=0,\omega=0}.
\label{eq:opo}
\end{eqnarray}
At the lowest nontrivial order the coupling between $\mu_H$ and $\omega$
is given by the term proportional to $c_{2,2}$ in Eq.~\eqref{eq:njk}.

The fluctuations of the helical density are described by $\chi_{\mu H}=-\partial^2\Omega_T/\partial\mu_H^2$;
from Eq.~\eqref{eq:njk} at $\mu_H\approx 0$ we read easily
\begin{equation}
\chi_{\mu H} = -c_{2,0} - \frac{c_{2,2}}{2}\omega^2.
\end{equation}
Within these approximations we can embed the effects of rotation on  $\chi_{\mu H}$ into the dimensionless coefficient 
$c_{2,2}$.

In the massless case $c_{2,0}$ and $c_{2,2}$ can be computed analytically with the result
\begin{eqnarray}
c_{2,0} &=& -\frac{N_c N_f}{3}T^2,\\
c_{2,2} &=& -\frac{N_c N_f}{2\pi^2}.\label{eq:chiHfinAAA}
\end{eqnarray}
Therefore in the massless limit we can write
\begin{equation}
\chi_{\mu H} = \frac{N_c N_f}{3}T^2 + \frac{N_c N_f}{4\pi^2}\omega^2.\label{eq:chiHfin}
\end{equation}
Note that the enhancement of fluctuations of helical density $\propto\omega^2$ in the above equation
is not related to the restoration of chiral symmetry, but to the fact that quark matter rotates. 
A similar rotation can be found for the susceptibility of the baryonic density.
 
\begin{figure}[t!]
  \centering
  % Requires \usepackage{graphicx
    \includegraphics[width=0.5\textwidth]{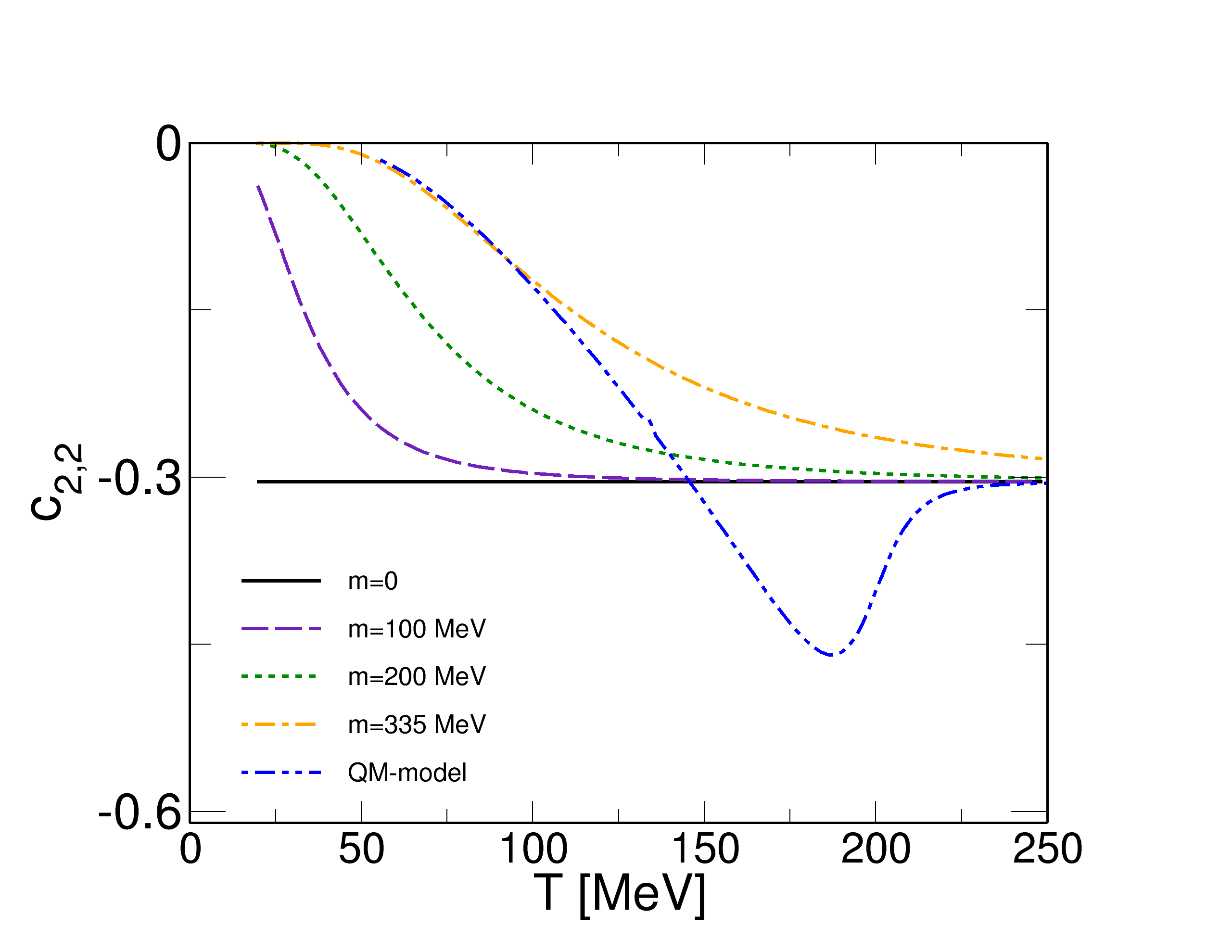}
  \caption{Coefficient $c_{2,2}$ versus temperature, for several values of the quark mass $m$ and for the QM model.}
  \label{Fig:c22}
\end{figure}

For $m\neq 0$
the calculation of $c_{2,2}$ has to be done numerically.  
We show $c_{2,2}$ versus temperature in Fig.~\ref{Fig:c22} for several values of   $m$.
Note that $c_{2,2}<0$ which means that   $\omega\neq 0$ enhances the fluctuations
of the helical density at a given temperature. This happens because $\omega$ modifies the single particle spectrum  
like a chemical potential, see Eq.~\eqref{eq:subracted444aaa}.

\subsection{The case of the QM model}

We limit ourselves to compute $\chi_H$ for $\mu=\mu_H=0$ to mimic the conditions 
of the medium produced in very high energy nuclear collisions. 
We write $\Omega=U + \Omega_q^\mathrm{ren} + \Omega_T$.
The thermal part of the thermodynamic potential
is given by Eq.~\eqref{eq:subracted444aaa} while the classical potential and the renormalized vacuum term 
are given by Eqs.~\eqref{Eq.EffectivePotential} and~\eqref{eq:renKLM} respectively.

We firstly show the coefficient $c_{2,2}$ in Fig.~\ref{Fig:c22}. This has been computed by numerical differentiation of
$\Omega$ similarly to Eq.~\eqref{eq:opo}
\begin{eqnarray}
c_{2,2} &=& \left.\frac{d^{4}\Omega}{d\mu_H^2 d\omega^2}\right|_{\mu_H=0,\omega=0}.
\label{eq:opo2}
\end{eqnarray}
At difference with Eq.~\eqref{eq:opo}, we have used the total derivative notation in Eq.~\eqref{eq:opo2} to emphasize
that the differentiation takes into account of the $\omega$ and $\mu_H-$dependence of the condensate.
$c_{2,2}$ is negative in agreement with the discussion for the ideal gas, signaling the enhancement of the fluctuations
induced by a nonzero $\omega$.
The most striking difference between the QM model and the fixed mass results is that in the former case the
chiral phase transition affects $c_{2,2}$, as it is clear from the groove structure around $T_c$.
Thus, in magnitude the enhancement of fluctuations due to $\omega\neq 0$ is larger than the one
obtained for an ideal gas.

\begin{figure}[t!]
  \centering
  % Requires \usepackage{graphicx
    \includegraphics[width=0.5\textwidth]{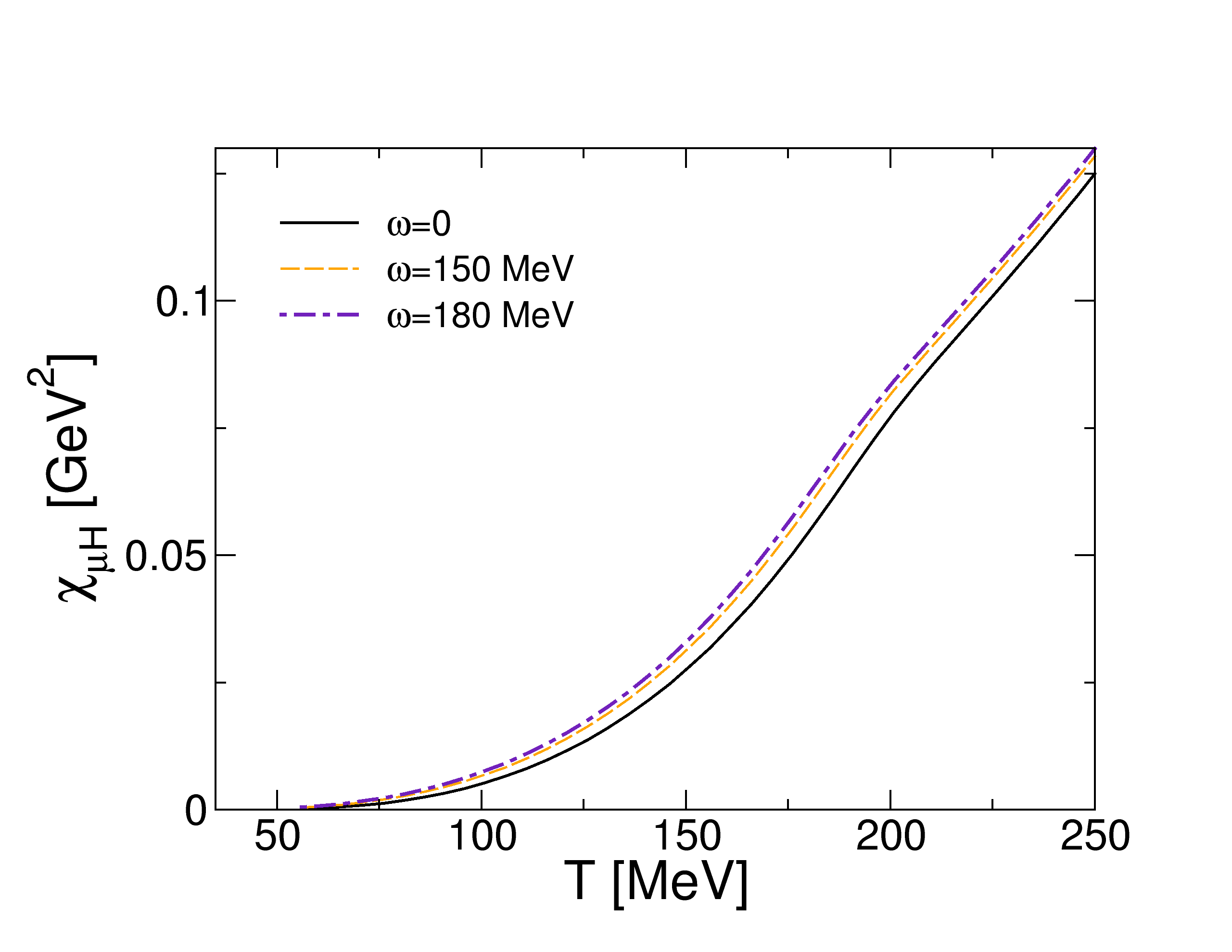}
  \caption{ $\chi_{\mu H}$ versus temperature, for several values of $\omega$ at $\mu_H=0$.}
  \label{Fig:susce22}
\end{figure}

We complete this study by computing  $\chi_{\mu H}$ versus temperature for $\omega\neq 0$ within the QM model.
The results are summarized in  Fig.~\ref{Fig:susce22}. The enhancement of the susceptibility induced by $\omega$ is evident.
We conclude that a rigid rotation enhances the fluctuations of helical density.

\section{Conclusions and Outlook}

We have studied the thermodynamics of quark matter with a helical chemical potential, $\mu_H$,
conjugated to helical density, $n_H$, together with the standard baryon chemical potential, $\mu$. 
We have analyzed chiral symmetry restoration as well as several susceptibilities of helical matter within 
the renormalized quark-meson model with two flavors of quarks. 
Helicity is a conserved quantity for free massive and massless quarks; moreover, 
its relaxation time in the hot quark-gluon plasma is larger than the lifetime of the
fireball produced in high energy nuclear collisions. For these reasons, it is meaningful to consider
the helical density, $n_H$, as a conserved quantity in heavy ion collisions.

We have computed the phase diagram at zero temperature: the result is summarized in Fig.~\ref{Fig:pdt0}.
We have found a critical line in the $\mu-\mu_H$ plane
which contains two critical endpoints. The middle portion of this line is made of a smooth crossover, while the 
lower right and upper left corners of the line are first order phase transitions. The symmetry of the phase diagram
around the line $\mu=\mu_H$ can be understood easily in terms of the duality of the thermodynamic potential
for the exchange $\mu\leftrightarrow\mu_H$. 
We have characterized the transition from the hadron gas at $T=0$ to the helical matter in terms of 
average helical and baryonic density, $n_B$; in particular, for helical matter $n_H\gg n_B$.
We have also considered the fluctuations
of helical and baryonic density. When the transition at large $\mu_H$ is considered, 
the fluctuations of $n_H$ are enhanced in comparison to those of $n$. This might be an additional difference,
besides $n_H \gg n_B$, between normal and helical quark matter.

We have then studied chiral symmetry restoration at finite temperature. The helical chemical potential
disfavors chiral symmetry breaking thus leads to a lower pseudo-critical line. We have completed the study by computing the
particle number susceptibilities  around the chiral crossover. Overall, our results support the idea that
$\mu_H$ makes the chiral phase transition softer and eventually for large values of the chemical potential,
the chiral critical endpoint disappears from the $T-\mu$ diagram.
Our results at zero as well as finite temperature are in agreement with \cite{Chernodub:2020yaf}.

We have also examined briefly the role of a rigid rotation on the fluctuations of helical density.
To mimic the conditions of the quark medium produced in very high energy nuclear collisions we have analyzed this problem
for the case $\mu=\mu_H=0$.
For the sake of simplicity, and since this is the first time in which this problem has been studied,
we have used the simplest implementation of the rotation of quark matter at finite temperature by modeling an unbounded system
in local equilibrium, limiting ourselves to study the region close to the rotation axis.
The susceptibility of the helical density, $\chi_{\mu H}$, is enhanced by the rotation.
We have illustrated this idea for an ideal gas, then analyzing the same problem within the
quark-meson model. Firstly, we have embedded the effect of rotation on $\chi_{\mu H}$ into a
dimensionless coefficient, $c_{2,2}$, shown in Fig.~\ref{Fig:c22} and computed analytically for the massless case,
see Eqs.~\eqref{eq:chiHfinAAA} and~\eqref{eq:chiHfin}.
Then we have done the computation of this coefficient for the quark-meson model,
finding that it is sensitive to the restoration of chiral symmetry at finite temperature.
Finally, for the QM model we have presented the full calculation of $\chi_{\mu H}$.
Our conclusion is that a rigid rotation enhances  the 
fluctuations of $n_H$ in a system in thermodynamic equilibrium.
Therefore, even though a medium with a net $n_H=0$ is formed in high energy nuclear collisions,
lumps of matter with $\langle n_H^2\rangle\neq 0$ can form because of event-by-event fluctuations, and this formation is favored
in rotating matter.
A more detailed study of rotation, including proper boundary conditions, will be the subject of a forthcoming article.

Altogether,
our result could have some impact on relativistic heavy ion collisions: in fact, 
lumps of matter with $\langle n_H^2\rangle\neq 0$ can be produced due to event-by-event fluctuations.
Thus, it is meaningful to question about the effect of helical density in the collisions.
We have found that helical density makes the chiral phase transition smoother
and lowers the critical temperature: 
it is likely that the local production of $n_H$ by fluctuations lowers the freezout temperature.
In addition to this, our model calculations show that the chiral phase transition 
is smoother in the lumps with $\langle n_H^2\rangle\neq 0$, and this
can affect the observables that are sensitive to the location of the critical endpoint in the $T-\mu$ plane.
While a complete study of the phenomenological impact of helical density on observables of heavy ion collisions
is well beyond the purpose of the present article,  
our results suggest that potentially helical matter can affect the evolution of the medium
created by  the collisions, and
we plan to report on this topic in the future.

\begin{acknowledgments}	
The authors acknowledge discussions with Roberto Anglani, Maxim Chernodub, Marco Frasca and Stefano Nicotri.
M. R. acknowledges John Petrucci for inspiration. 
The work of the authors is supported by the National Science Foundation of China (Grants No.11805087 and No. 11875153).
\end{acknowledgments}

\end{document}